\documentclass[12pt]{article}
\usepackage{epsfig}
\usepackage{latexsym}
\begin{document}
\title{On the Gravitational Wave in de Sitter Spacetime\footnote{Supported by the National Natural Science Foundation of China under Grant No. 10373003}}
\author{\begin{tabular}{c}\bigskip Liu Liao$^{}$\footnote{Email: liuliao1928@yahoo.com.cn } \\  \smallskip Department of Physics, Beijing Normal
University, Beijing, 100875, PRC
\end{tabular}}
\maketitle

\begin{abstract}
For there is always a wrong sign in the mass of graviton in the
so-called perturbation expansion approximation of both Minkowski
and de Sitter spacetimes, the existence of gravitational wave from
the metric perturbation of de Sitter spacetime is doubtful. We try
another way to start from the assumption that the gravitational
wave equation should be both general covariant and conformal
invariant and find that graviton is no
longer a part of metric field, it has an effective mass of $m_g=\sqrt{R/6}=%
\sqrt{2\Lambda/3}$ with correct sign in de Sitter spacetime,
though it's intrinsic mass remains zero.
       \\
       \\
PACS number(s):  03.50.Kt, 04.20.Cv
\end{abstract}

As was shown by S.Perlmutter et al in 1999 \cite{S}, that a best
fit flat cosmology to the observed Hubble constant $H_0$ and  cosmological constant $%
\Lambda$ from the red-shift of $I_a$-type supernova can be obtained,  that
is
\begin{eqnarray}
\Omega_m+\Omega_\Lambda&=&1  \nonumber \\
(\Omega_m,\Omega_\Lambda)&=&(0.28,0.72)
\end{eqnarray}
wherein
\begin{eqnarray}
\Omega_{m,_ \Lambda}&\equiv&\frac{\rho_{m,\Lambda}}{\rho_c}, \rho_c\equiv%
\frac{3H_0^2}{8\pi G}  \nonumber
\end{eqnarray}
The above result suggests strongly that not only the so-called
 ``cosmological vacuum energy density $ \frac{\Lambda}{8\pi
G}=\rho_\Lambda $" or ``dark energy" is about 2.6 times that of
the matter field in our universe but also we are living in an
accelerating expanding universe. Now a very fundamental problem
is, does the gravitational wave exist in the de Sitter universe?

First we try to introduce the linear approximation of perturbation
expansion.

One of the perturbation expansion initiated by C.M$\phi$ller \cite%
{C} is to expand the spacetime metric $g_{\mu\nu}$ around the
Minkowski spacetime metric $\eta_{\mu\nu}$ as follows
\begin{equation}
g_{\mu\nu}=\eta_{\mu\nu}+h_{\mu\nu}
\end{equation}
where
\begin{eqnarray}
(\eta)=(+1,-1,-1,-1)  \nonumber
\end{eqnarray}
We shall show later, this kind of perturbation expansion can lead
us to the finite-range gravitation.  Note, throughout this paper,
the signature convention is the same as that of Birrell and
Davies' book \cite{N}. It is straightforward to show that  the
linear approximation of Einstein's gravitational field equation
with cosmological term
\begin{equation}
R_{\mu\nu}-\frac{1}{2}Rg_{\mu\nu}+\Lambda g_{\mu\nu}=-8\pi GT_{\mu\nu}^{(m)}
\end{equation}
is \cite{C}
\begin{equation}
(\Box-2\Lambda)h_{\mu\nu}=-16\pi GT_{\mu\nu}
\end{equation}
where $R_{\mu\nu}=\frac{1}{2}\Box h_{\mu\nu}$ , in harmonic gauge
\begin{eqnarray}
\Box&\equiv&\eta^{\mu\nu}\partial_\mu\partial_\nu  \nonumber \\
T_{\mu\nu}&\equiv&\Gamma_{\mu\nu}^{(m)}+T_{\mu\nu}^{(\Lambda)}\equiv
T_{\mu\nu}^{(m)}-\frac{1}{2}T^{(m)}\eta_{\mu\nu}-\frac{\Lambda}{8\pi G}%
\eta_{\mu\nu}  \nonumber
\end{eqnarray}
Note, the correct mass term for $h_{\mu\nu}$ in Eq.(4) should be
positive in the above signature, that is the mass
$m_g^2=(-2\Lambda)$ should be positive or $\Lambda$ should be
negative! (See also the paper of Freund et al in \cite{P})

Therefore Eq.(4) for the perturbed gravitational field $h_{\mu\nu}$ seems to
have a mass term only when $\Lambda<0$, in other words, the perturbed
gravitational field $h_{\mu\nu}$ in the above approximation can be looked
upon as a spin 2 particle with rest mass $m_g=\sqrt{-2\Lambda}$  $(\sqrt{%
-2\Lambda}\hbar c^{-1})$ in flat Minkowski spacetime background when $%
\Lambda<0$ .

After introducing
\begin{eqnarray}
\hat{\rho}&=&\rho_m-\frac{\Lambda}{4\pi G}  \nonumber \\
g_{00}&=&1+2\varphi
\end{eqnarray}
in static case, Eq.(4) gives the famous Newmann-Seeliger
gravitational field equation when $\Lambda<0$
\begin{eqnarray}
\Delta\varphi+2\Lambda\varphi=4\pi G\hat{\rho}  \nonumber
\end{eqnarray}
or
\begin{eqnarray}
\Delta\varphi-(-2\Lambda)\varphi=4\pi G\hat{\rho}
\end{eqnarray}
the above mass term implies there exists an Yukawa-type potential
\cite{M}
\begin{eqnarray}
\varphi=-\frac{GM}{r}e^{-\sqrt{-2\Lambda}\cdot r}
\end{eqnarray}

However on account of the smallness of $|\Lambda|$ or of graviton
mass, the deviation from Newton's law has no practical measurable
meaning inside the Hubble length
$H^{-1}=\sqrt{-3/\Lambda}\simeq10^{28}$ cm of our universe.

For the Minkowski spacetime is no longer a solution of the
Einstein's gravitational field equation with $\Lambda\neq0$, many
authors argued \cite{e} that the perturbation expansion  of
$g_{\mu\nu}$ should around the de Sitter spacetime. Now let
$\hat{g}_{\mu\nu}$ be the metric of de Sitter spacetime, the
perturbation expansion of certain metric $g_{\mu\nu}$ around it
wil be
\begin{eqnarray}
g_{\mu\nu}=\hat{g}_{\mu\nu}+h_{\mu\nu}
\end{eqnarray}
or
\begin{eqnarray}
\delta g_{\mu\nu}=h_{\mu\nu}\nonumber
\end{eqnarray}
then the perturbed Einstein vacuum field equation with
$\Lambda$-term
\begin{eqnarray}
G_{\mu\nu}+\Lambda g_{\mu\nu}=0
\end{eqnarray}
will be
\begin{eqnarray}
\delta G_{\mu\nu}+\Lambda h_{\mu\nu}=0
\end{eqnarray}
Novello and Neves showed \cite{P} that the field equation for the
perturbation (10) should be
\begin{eqnarray}
\Box h_{\mu\nu}-\frac{2}{3}\Lambda h_{\mu\nu}=0\
\end{eqnarray}
wherein gauge conditions
\begin{eqnarray}
h_\mu^\mu=0 ,  h_{\nu;\mu}^\mu=0
\end{eqnarray}
are used in order that the field $h_{\mu\nu}$ has just five
degrees of freedom. From Eq.(11), it's evident that the mass term
of $h_{\mu\nu}$ has a wrong negative sign or the graviton should
have an imaginary mass
\begin{eqnarray}
m_g=\sqrt{-2\Lambda/3}
\end{eqnarray}
The perturbation field equations Eq.(4) and Eq.(11) have meaning
only for anti-de Sitter spacetime of $\Lambda<0$. Unfortunately,
the accelerating expanding of our universe strongly suggests that
the cosmological constant $\Lambda$ is positive, this may imply
that the above two different kinds of perturbation expansion are
not a reasonable approximation when $\Lambda>0$.

It seems therefore the metric perturbation of both Minkowski
universe and de Sitter universe can't create gravitational wave or
graviton when $\Lambda>0$. However, people believe that violent
astronomical events, e.g., the burst of supernova, collision of
galaxies or double stars damping... certainly should radiate
gravitational wave. But if the above cited violent astronomical
events can be looked upon as perturbations to the metric of de
Sitter universe for the cosmological distant observers, then our
above citation will cast strong doubt on the existence of such
gravitational wave or graviton. So all gravitational wave
detecting projects based on the metric perturbation of de Sitter
universe might have no results. This may really be a very severe
problem confront us !

In what follows we try another way to manipulate the gravitational
wave problem in de Sitter spacetime. There are evidences to
suppose that our universe may be approximated by the steady state
model
\begin{eqnarray}
ds^2=dt^2-e^{2Ht}(dr^2+r^2d\theta^2+r^2\sin^2\theta d\varphi^2)
\end{eqnarray}
where $H=\sqrt{\Lambda/3}$

Though this model is not flat, it is however both spacial flat and
conformal flat to the Minkowski spacetime \cite{N}, this implies
that there is a conformal factor $\Omega ^{2}$ such that
\begin{eqnarray}
ds^{2}=\Omega ^{2}ds_{M}^{2}
\end{eqnarray}%
where
\begin{eqnarray}
ds_{M}^{2} &=&\eta _{\mu \nu }dx^{\mu }dx^{\nu }  \nonumber\\
\Omega ^{2} &=&e^{2Ht}  \nonumber
\end{eqnarray}%
We can first construct a Lorentzian covariant massless tensor
field theory in Minkowski spacetime, then by using the conformal
relation Eq.(15) we can construct a conformal invariant and
general covariant tensor field theory in the curved de Sitter
spacetime as follows :

The Lorentzian covariant massless free tensor field equation in
Minkowski spacetime reads
\begin{eqnarray}
\Box_\eta\varphi_{\mu\nu}^{(M)}=0
\end{eqnarray}

We remark a self-consistent tensor field equation can be established, the
details can be seen from the supplementary remarks in Wentzel's book \cite{G}%
. By conformal transformation, a conformal invariant and general
covariant tensor field $\varphi_{\mu\nu}$ equation in conformal
flat de Sitter spacetime is \cite{N}
\begin{eqnarray}
[\Box+\frac{1}{4}(n-2)R/(n-1)]\varphi_{\mu\nu}=0  \nonumber
\end{eqnarray}
or
\begin{eqnarray}
\Box\varphi_{\mu\nu}+\frac{2}{3}\Lambda\varphi_{\mu\nu}=0\
\end{eqnarray}
where the dimension of space time is $n=4$, the curvature scalar R
of de Sitter spacetime is $R=4\Lambda$
\begin{eqnarray}
\varphi_{\mu\nu}=\Omega^{-1}\varphi_{\mu\nu}^{(M)}
\end{eqnarray}

I would like to point out that the general covariance of Eq.(17)
can be guaranteed by choosing the harmonic gauge
\begin{eqnarray}
\Box ' (\alpha_\mu^{' \lambda})=\frac{\partial}{\partial
x^{\mu}}(\Box ' x^{' \lambda})=0\nonumber
\end{eqnarray}

Here we emphasize that the conformal invariance assures that the
intrinsic mass of $\varphi _{\mu \nu }^{(M)}$ and $\varphi _{\mu
\nu }$ are always zero, and its effective mass $m_{g}\equiv
\sqrt{2\Lambda /3}$ comes wholly from the non-zero curvature
scalar R or cosmological constant $\Lambda $, also remark $\varphi
_{\mu \nu }$ and $\varphi _{\mu \nu }^{(M)}$ in Eq.(18) are
conformal equivalent, they should have the same Fock
representation, so
particle picture or graviton has meaning for both $\varphi _{\mu \nu }$ and $%
\varphi _{\mu \nu }^{(M)}$, though it is not true in general,
especially in the case $\Lambda <0$ (anti-de Sitter spacetime). It
is interesting to point out that the effective mass in Eq.(17) is
$m_{g}=\sqrt{2\Lambda /3}$, which is very similar to  Novello's
result ($m=\sqrt{-2\Lambda /3}$) except that our result has the
correct positive sign in de Sitter background spacetime.

Let's give a brief summary of our new approach. First of all, no
perturbation expansion of any kind is used throughout our paper.
Therefore unlike the perturbation expansion, our tensor field is
not the perturbed part of the background metric field. Now if the
self-consistent massless tensor field in Eq.(16) can be looked
upon as graviton in Minkowski spacetime, we have then no other
choice except to receive the interpretation that it's conformal
invariant partner (i.e., conformal coupling partner) should be
looked upon as conformal graviton in de Sitter spacetime with
non-zero effective rest mass of $\sqrt{2\Lambda/3}$.

Our last conclusion is therefore.

It seems, there are no gravitational wave or graviton originated
from metric perturbation of de Sitter universe, but conformal
gravitational wave or conformal graviton with effective mass
$\sqrt{2\Lambda/3}$ may exist.

\section*{Acknowledgement}

I am grateful to the discussion and e-communication with Profs.
 S.Deser, C.G.Huang, M.Novello, S.Y.Pei and
Dr. J.B.Pitts. Also, I would like to thank Dr. J.S.Yang for his
painstaking help to this manuscript.

\end{document}